\documentclass[12pt]{article}

\usepackage{amsfonts}

\begin{document}



\def\Bid{{\mathchoice {\rm {1\mskip-4.5mu l}} {\rm
{1\mskip-4.5mu l}} {\rm {1\mskip-3.8mu l}} {\rm {1\mskip-4.3mu l}}}}


\newcommand{\beq}{\begin{equation}}
\newcommand{\eeq}{\end{equation}}
\newcommand{\bea}{\begin{eqnarray}}
\newcommand{\eea}{\end{eqnarray}}
\newcommand{\eL}{{\cal L}}
\newcommand{\half}{\frac{1}{2}}
\newcommand{\J}{\bf J}
\newcommand{\bP}{\bf P}
\newcommand{\G}{\bf G}
\newcommand{\K}{\bf K}
\newcommand{\M}{{\cal M}}
\newcommand{\bu}{\bf u}
\newcommand{\la}{\lambda}
\newcommand{\ohalf}{\textstyle{1 \over 2}}
\newcommand{\thd}{\textstyle{1 \over 3}}
\newcommand{\tthd}{\textstyle{2 \over 3}}

\begin{flushright}
DOE-ER-40757-123\\
UTEXAS-HEP-99-1
\end{flushright}

\hfil\hfil

\begin{center}
{\Large {\bf Clifford Periodicity from Finite Groups}}
\end{center}
\begin{center}
\today
\end{center}
\hfil\break
\begin{center}
{\bf Luis J. Boya \footnote{Permanent address  Departamento de F{\'{\i}}sica 
Teorica, Facultad de Ciencias.  Universidad de Zaragoza, E-50009 Zaragoza, 
Spain.  Email:  luisjo@posta.unizar.es} and 
Mark Byrd \footnote{mbyrd@physics.utexas.edu}\\}
\hfil\break
{\it Center for Particle Physics \\
University of Texas at Austin \\
Austin, Texas 78712-1081}
\end{center}
\hfil\break

\begin{abstract}
We deduce the periodicity 8 for the type of $Pin$ and $Spin$ representations  
of the orthogonal groups $O(n)$ from simple combinatorial properties of the 
finite Clifford groups generated by the gamma matrices.  We also include the 
case of arbitrary signature $O(p,q)$.  The changes in the type of 
representation can be seen as a rotation in the complex plane.  The essential 
result is that adding a $(+)$ dimension performs a rotation by $\pi/4$ in 
the counter clock-wise sense, but for each $(-)$ sign in the metric, the 
rotation is clockwise.
\end{abstract}


\section{Introduction}

The periodicity of Clifford algebras, first described by Atiyah, Bott and 
Shapiro \cite{abs},
$$
C_{n+8} = C_n \otimes C_8,
$$
is a fundamental mathematical discovery.  It is related to Bott's 
periodicity of the homotopy groups of the classical groups.  It is 
essential in $K$-theory, in the solution by Adams of the vector field 
problem in spheres, {\it etc.}  It is our aim to give a short proof 
of this important but simple property.

In this Note we shall obtain this periodicity from elementary properties 
of the representations of finite groups, namely the multiplicative groups 
generated by the ``Dirac gamma matrices'' of the Clifford algebras.  This 
finite group has been considered in the past \cite{adams}, but 
not to our knowledge, applied to this problem.

If we have a positive quadratic form over the reals with isometry group $O(n)$, recall that the Clifford algebra is obtained by linearizing it, \`{a} la Dirac: 
$$
\left(\sum x^\mu\right)^2 = \left(\sum x^\mu \gamma_\mu\right)^2.  
$$

The algebra is generated by the $\gamma$'s, where
\begin{equation}
\{\gamma_{\mu},\gamma_{\nu}\} = 2 \delta_{\mu\nu}.
\label{comrel}
\end{equation}

There is a {\sl finite} multiplicative group
\begin{equation}
\Gamma = \{\pm \Bid, \pm \gamma_\mu, \pm \gamma_\mu\gamma_
\nu, ... , \pm \gamma_1\dots\gamma_n \equiv \pm\gamma_{n+1}\},
\label{gamma}
\end{equation}
with $\gamma_{n+1} = \Pi_n \gamma_\nu$, which generates the whole $Pin(n)$ 
group, and the even part
$$
\Gamma_0 = \{\pm \Bid,\pm \gamma_\mu\gamma_\nu, \pm \gamma_\lambda 
\gamma_\mu \gamma_\nu \gamma_\rho, \dots \},
$$
which generates $Spin(n)$.

For our purposes we shall need two well-known results from the representation 
theory of a finite group $G$ (see {\it eg.} Bacry \cite{Bacry} or Br\"ocker 
and Dieck \cite{bandd}).

{\bf (A) Burnside Theorem:}  {\it The group algebra is a direct sum of complete 
matrix algebras, and there are as many of these as there are classes of 
conjugate elements in $G$.  Let $|G|$ be the order of $G$,}
\begin{equation}
|G| = \sum_{\mbox{\scriptsize{classes}}} (d_i)^2,
\label{Burn}
\end{equation}
{\it where $i$ runs through the irreducible representations (irreps) of $G$, the 
same as the number of classes, and $d_i$ is the dimension of the 
$i^{\mbox{th}}$ irrep.}

Both $\Gamma$ and $\Gamma_0$ are nearly Abelian in the sense that the commutator 
subgroup is very small.  Hence the abelianized quotient group is very large.  
Most of the irreps, in fact all except one or two, are therefore one dimensional.

{\bf (B)  The type}, $i$, {\it of a particular representation} $D$ {\it is given by the expression}
\begin{eqnarray}
i(D) &=& \frac{1}{|G|}\sum_g \chi(g^2) \nonumber \\
              &=&  \left\{ \begin{array}{lll}
                                +1 &  \hbox{for real irreps},\\
                     \phantom{+} 0 &  \hbox{for complex irreps},  \\
                                -1 &  \hbox{for quaternionic, q-real,
                                             or quasireal irreps} 
                        \end{array} \right.
\label{type}
\end{eqnarray}
{\it where} $\chi_D(g) = \mbox{Tr}(D(g))$ {\it is the character of element $g$ in the representation $D$.  The result} (\ref{type}) {\it holds also for compact groups} \cite{bandd}.

In our case the sums in (\ref{type}) are easy to compute because the anticommutativity (\ref{comrel}) implies the squares $(\pm \gamma_\lambda \gamma_\mu \gamma_\nu \gamma_\rho \dots)^2 = \pm \Bid$, and the number of them is a simple combinatorial number.


\section{Periodicity for $Pin(n)$}


\subsection{Even Dimension}

First let $n=2\nu$ be even.  We have $|\Gamma|= 2^{n+1} = 2^{2\nu+1}$.  The 
commutator $[\Gamma,\Gamma]$ is clearly $= \mathbb{Z}_2$.  Hence the 
abelianized quotient is half as large as $\Gamma$:
$$
|\Gamma/[\Gamma,\Gamma]| = 2^{2\nu}.
$$

There are also $2^{2\nu}+1$ conjugation classes (all binomial terms plus one, 
which is the only nontrivial central), so there is a {\sl unique} solution 
to Burnside's numerical equation (\ref{Burn})
$$
2^{2\nu+1} = 2^{2\nu} \cdot 1^2 +1\cdot(2^\nu)^2,
$$
and there is a single irrep of dimension $2^\nu$, the {\it (s)pin} representation, $\Delta$.

The {\sl type} of $\Delta$ is easy to compute; it is
$$
i(\Delta) = \frac{2 \cdot 2^\nu}{2^{2\nu+1}}\left[1+{2\nu \choose 1}-{2\nu \choose 2}-{2\nu \choose 3} + {2\nu \choose 4} +\cdots\right]
$$
The factor of $2$ in the numerator is the $\pm$ sign in (\ref{gamma}) and 
$2^\nu$ is the dimension of $\Delta$.  It is clear why the signs alternate 
in blocks of two:  $\gamma_\nu^2 = +1$ implies 
$(\gamma_\mu \gamma_\nu)^2 = -1$ which implies 
$(\gamma_\mu \gamma_\nu \gamma_\rho)^2 = -1$, {\it etc.}  Hence,
\begin{eqnarray}
i(\Delta) &=& \frac{1}{2^\nu}\left[1 -{2\nu \choose 2} + {2\nu \choose 4} -+\cdots\right] 
          + \frac{1}{2^\nu}\left[{2\nu \choose 1} - {2\nu \choose 3} +- \cdots\right]
\nonumber \\
          &=& \frac{1}{2^\nu}[\mbox{Re}(1+i)^{2\nu} + \mbox{Im}(1+i)^{2\nu}] = 
(\mbox{Re} + \mbox{Im})[(1+i)/\sqrt{2}]^{2\nu}, \nonumber
\end{eqnarray}
which gives
\begin{equation}
i(\Delta) = (\mbox{Re} + \mbox{Im})e^{2\pi i n/8}
\label{evenper}
\end{equation}
for $n = 2 \nu$, even.  So that for $n$ even the periodicity is clearly seen 
to be eight,
$$
i(\Delta) = \cos(2\pi n/8) + \sin(2\pi n/8).
$$
Notice that there are {\it no} complex irreps for the $Pin(2\nu)$ groups.


\subsection{Odd Dimension}

The computation for $n = 2\nu +1$ odd is similar.  
$$
|\Gamma| = 2\cdot 2^{2\nu + 1} = 2^{2\nu +2}
$$
The Burnside relation gives
$$
2^{2\nu +2} = 2^{2\nu + 1}\cdot 1^2 + 2 \cdot (2^{2\nu})^2.
$$
There are now {\it two} $Pin(2\nu +1)$ irreps of the same type; call them still 
$\Delta$.  
\begin{eqnarray}
i(\Delta) &=& \frac{2\cdot2^\nu}{2^{2\nu +2}}\left[1+{2\nu + 1 \choose 1}
-{2\nu + 1 \choose 2}-{2\nu+ 1 \choose 3} + \cdots\right] \nonumber \\
          &=& \frac{1}{\sqrt{2}}(\mbox{Re} + \mbox{Im})[(1+i)/\sqrt{2}]^{2\nu+1}
\end{eqnarray}
so that
\begin{equation}
i(\Delta) = \frac{1}{\sqrt{2}}(\cos(2\pi n/8) + \sin(2\pi n/8))
\end{equation}
for $n = 2\nu +1$ odd.  This, together with (\ref{evenper}), completes 
the periodicity 8:
\begin{eqnarray}
i(\Delta) &=& 1,1,1,0,-1,-1,-1,0,1,1,1,\dots \nonumber \\
       n  &=& 0,1,2,3,\phantom{+}4,\phantom{+}5,\phantom{+}6,7,8,9,10, \dots \nonumber
\end{eqnarray}
The essential, simple result is that adding a dimension (in this ``Euclidean'' case) corresponds to a rotation of $\pi/4$.

\section{Periodicity for $Spin(n)$}

Now we use the restricted finite Clifford group
$$
\Gamma_0 = \{ \pm \Bid, \pm \gamma_\mu \gamma_\nu, 
\pm \gamma_\lambda \gamma_\mu\gamma_\nu\gamma_\rho, \dots \},
$$
$$
|\Gamma_0| = 2^n
$$
Let $n = 2\nu$ even.  The Burnside relation gives 
$$
2^{2\nu} = 2^{2\nu-1}\cdot 1^2 +2\cdot (2^{\nu-1})^2.
$$
The two spin irreps are the traditional $\Delta^\pm$.  So for $n$ even their 
types are given by
\begin{eqnarray}
i(\Delta^\pm) &=& \frac{2\cdot2^{\nu-1}}{2^{2\nu}}
\left[1-{2\nu \choose 2}+{2\nu \choose 4} -\cdots \right] \nonumber \\
              &=& \cos(2\pi n/8).
\label{evenper2}
\end{eqnarray}

For $n = 2\nu+1$ odd, the Burnside relation gives 
$$
2^{2\nu + 1} = 2^{2\nu} \cdot 1^2 + 1\cdot (2^\nu)^2.
$$
The type of the representation is then
\begin{equation}
i(\Delta^+) = \sqrt{2} \cos(2\pi n/8)
\label{oddper2}
\end{equation}
for $n$ odd.  
Combining (\ref{evenper2}) and (\ref{oddper2}) we recover the usual $Spin(n)$ 
periodicity 8:
\bea
i(\Delta^{\pm}) &=& 1,1,0,-1,-1,-1,0,1,1,1,\;0 \cdots \nonumber \\
      n   &=& 0,1,2,\phantom{+}3,\phantom{+}4,\phantom{+}5,6,7,8,9,10,  \cdots \nonumber
\eea
The relation between $Pin(n-1)$ with $Spin(n)$ is to be expected since the 
corresponding complete Clifford algebras coincide \cite{abs}.


\section{The Case of Signature}

It is easy to extend the results above for a metric with signature $(p,q)$ 
where $p,q$ are {\sl arbitrary} positive integers.  Now we have more groups:
$$
 O(p,q), \;\;\;\;\;\; SO(p,q), \;\;\;\;\;\; SO_0(p,q)
$$
where the $SO_0(p,q)$ is the connected part.  Now the finite group $\Gamma$ generates
 $Pin(p,q)$, but the restricted group, $\Gamma_0$, generates only 
$Spin(p,q)$, which covers $SO_0(p,q)$ twice.

The signature complication is inessential, as {\sl each negative sign dimension 
corresponds 
to a $\pi/4$ rotation in the opposite (clockwise) sense}.  To prove this, it 
is enough to reckon the type for the negative-definite metric, $(0,n)$.  Now
  $(\gamma_\mu)^2 = -1$, so sets of odd numbers of $\gamma$'s change sign, 
but the even sets do not.  Hence,
\begin{equation}
\mbox{Type}(0,n) = {\cal P}\left(\frac{(1-i)}{\sqrt{2}}\right)^n 
= {\cal P}\exp(-2\pi i n/8)
\label{sigtype}
\end{equation}
where ${\cal P}$ is the projection (with the appropriate factor of $\sqrt{2}$ 
as before), (Re +Im) for the complete $Pin$ group, and Re only for the 
$Spin$ part.

As the angles add indepedently, we have 
\begin{eqnarray}
\mbox{Type}(p,q) &=& {\cal P}[\exp(2\pi i p/8)\exp (-2\pi i q/8)] \nonumber \\
                 &=& {\cal P}[\exp(2\pi i (p-q)/8)
\label{typepq}
\end{eqnarray}
which of course can be proved directly from the sums
$$
\left(1 + {p-q \choose 1} -  {p-q \choose 2} -{p-q \choose 3} + {p-q \choose 4}
 \cdots  \right)
$$

$$
\left(1 + {p \choose 1} -  {p \choose 2} - {p \choose 3} 
 \cdots  \right)\left(1 - {q \choose 1} -  {q \choose 2} + {q \choose 3} 
 \cdots  \right).
$$
We have that $Pin(p,q) \neq Pin(q,p)$, but the $Spin$ groups are the same.  The 
double covering of the connected part is unique, but the extensions from 
$O(p,q)$ and $O(q,p)$ are different.

Formula (\ref{typepq}) is our final result.  It shows 8-periodicity in the 
signature $(p-q)$, which is well-known.  We recall some consequences.

\begin{itemize}
\item{The so-called split forms $(p,p)$ and $(p+1,p)$ are real.}
\item{The Lorentzian metric $(p,1)$ has type $(p-1)$, so it is 2 in Minkowski 
space, regardless of whether it is $(3,1)$ or the light cone $(2,0)$.}
\item{The same for the conformal extension $O(p,q) \rightarrow O(p+1,q+1)$, 
the type is still that of $(p,q)$.}
\item{The Lorentz groups $O(25,1)$ and $O(9,1)$ used in string theory are 
of the real type.  This, no doubt, is crucial for the scale anomaly 
cancellation.}
\item{The anomaly free gauge group $O(32)$ used in Type I and Heterotic string 
theory is also of the real type.}
\end{itemize}


\section{Final Remarks}

Clifford periodicity 8 for the real orthogonal groups is an important 
phenomenon; so we find it satisfying to be able to provide a proof that 
is intrinsic, {\it i.e.}, does not depend on the particular representation 
of the gamma matrices.  It also covers the $Pin$ as well as the $Spin$ 
groups, and deals with the case of arbitrary signature.

The existence of two groups for the full orthogonal group has found an 
interesting application in the paper \cite{cecile2}.  In fact, the 
reflection properties of spinors {\it do} depend on the sign of the 
metric, and even in the ``skeleton'' finite Clifford group this difference 
shows up.

We might mention another periodicity shown by one of us \cite{ljb}, which 
should be related to the case discussed here; namely the optical theorem in 
quantum mechanical scattering.  This depends also on the dimension of the 
space with periodicity 8 (although it has other factors, such as the volume 
of the sphere and the inverse of the momentum to some power, that depend on 
the dimension of the space as well).  The formula reads \cite{ljb}
\begin{equation}
\sigma_{tot} + 2\left(\frac{2\pi}{k}\right)^{(n-1)/2}
\mbox{Re}\{e^{2\pi i(n-1)/8} f(0)\} = 0 
\end{equation}
where $\sigma_{tot}$ is the total elastic scattering cross section in 
$n$-dimensional space and $f(0)$ is the forward scattering amplitude.  The 
similarity with the results above is striking and the reason, we think, is the
 same:  the wave function is a kind of ``square root'' of an orthogonal 
observable, and hence behaves like a spinor.  This argument was already 
advanced in \cite{ljb}.

Finally we call attention to the book \cite{bandt} in which there is 
also a ``clock'' with $\mathbb{Z}/8$ rotations.


\section*{Acknowledgements}

LJB would like to thank the Center for Particle Physics and in particular 
Prof. E. C. G. Sudarshan for partial support, and to the spanish DGAICYT, 
grant \#AEN-97-1680.  MSB would like to thank DOE for its support under 
the grant number DOE-ER-40757-123.


\end{document}